\documentclass[preprint,showpacs,showkeys,preprintnumbers,aps]{revtex4}

\usepackage{graphicx}

\begin{document}

\preprint{Balke et al., CoTi$_{1-x}$Fe$_x$Sb.}

\title{A new diluted magnetic semiconductor: \\
                The half-metallic ferromagnet CoTi$_{1-x}$Fe$_x$Sb.}

\author{Benjamin Balke, Kristian Kroth, Gerhard H. Fecher,  and Claudia Felser}
\email{felser@uni-mainz.de}
\affiliation{Institut f\"ur Anorganische und Analytische Chemie, \\
Johannes Gutenberg - Universit\"at, D-55099 Mainz, Germany.}

\date{\today}

\begin{abstract}

Half-Heusler compounds with 18 valence electrons are semi-conducting.
It will be shown that doping with electrons results in half-metallic
ferromagnets, similar to the
case of diluted semi-conductors.
CoTiSb is known to be a semi-conducting Half-Heusler compound. Doping by
Fe is expected to result in ferromagnetic order. It was found that Ti
can be replaced by up to about 10\% Fe while its crystal structure
still remains $C1_b$, which was proved by X-ray powder diffraction.
SQUID magnetometry revealed a magnetic moment of 0.32$\mu_B$ per unit
cell at 5K.
\end{abstract}



\maketitle

\section{Introduction}
\label{IN} Within the last decade a new idea has revolutionised
electronic devices: To use the spin of electrons in addition to its
charge. This field of application is called spintronics \cite{Coe01}.
Half-metallic ferromagnets (HMF) seem to be a suitable class of
materials which meet all requirements of spintronics. One reason is
their exceptional electronic structure: They are metals in one spin
direction and semiconductors in the other. The Half-Heusler compound
NiMnSb was the first material being predicted by electronic structure
calculations to be a HMF \cite{GME83}.

Many attempts have been made to prepare semi-conducting compounds which
also have ferromagnetic properties. Mn-doped GaAs was considered to be
a suitable compound, but its Curie temperature is only about 150~K
\cite{EWC02} and thus still far away from being suitable for
application in electronic devices (see \cite{Mac05} for a recent
review). Heusler compounds exhibit a large variety of different
electronic and magnetic properties ranging from semi-conducting to
ferromagnetic. This work focuses on the search for suitable compounds
that bridge both semi-conducting and ferromagnetic properties. In
particular, the research concentrates on finding half-metallic
ferromagnetic Half-Heusler alloys with transition temperatures well
above room temperature.

\section{Results and discussion}

CoTi$_{1-x}$Fe$_x$Sb samples were prepared by arc melting of
stoichiometric amounts of the constituents in an argon atmosphere at
10$^{-4}$~mbar. The polycrystalline ingots were then annealed in an
evacuated quartz tube at 1000~K for 7~days.

\subsection{Electronic properties}

The electronic structure was calculated for pure and Fe doped CoTiSb in
order to examine their electronic and magnetic structure. Self
consistent calculations were performed by means of the
full-relativistic, spin polarised Korringa-Kohn-Rostocker (KKR) method
in combination with coherent potential approximation (CPA)
\cite{Ebe99}. As predicted by Tobola {\it et al} \cite{Tob00}, CoTiSb
turned out to be a semiconductor with a gap at the Fermi energy for
both spin directions. Partial replacement of titanium by iron (10\%)
(assuming that Fe is statistically distributed on the Ti positions),
has the result that the semiconductor is converted into a HMF: The DOS
at the Fermi energy is clearly different from zero for only one spin
direction, while remaining zero for the other. Measurements of the
conductivity revealed the metallic character of the Fe-doped samples in
contrast to the semi-conducting behaviour of the pure sample.

\subsection{Structural properties}

%%%%%%%%%%%%%%%%%%%% Figure 1 %%%%%%%%%%%%%%%%%%%%%%%%%%%%%%%%%%%%%%%%%%%%%%%%
  \begin{figure}[t]
\includegraphics[scale =1.0]{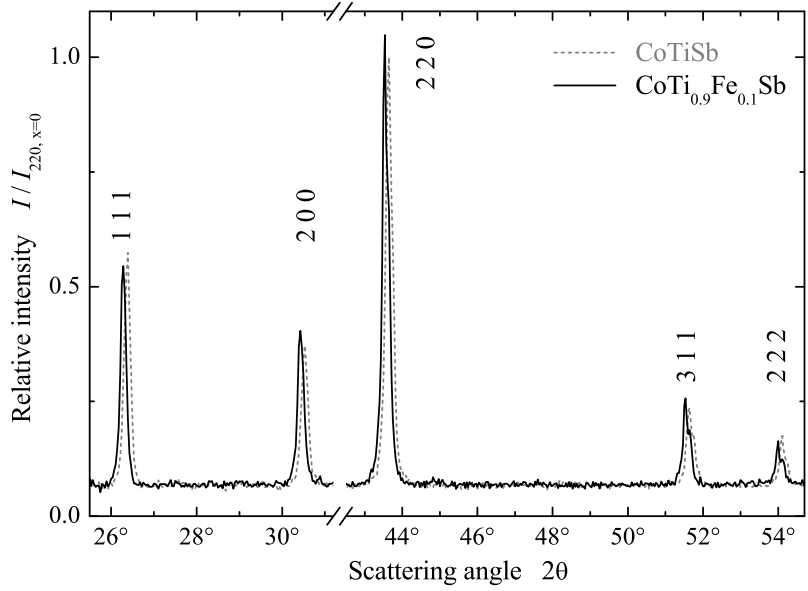}
\centering
\caption{X-ray diffraction pattern of CoTiSb and CoTi$_{0.9}$Fe$_{0.1}$Sb.
        The spectra were excited by Cu K$_\alpha$ radiation}
    \label{fig-1}
  \end{figure}
%%%%%%%%%%%%%%%%%%%%%%%%%%%%%%%%%%%%%%%%%%%%%%%%%%%%%%%%%%%%%%%%%%%%%%%%%%%%%%

It was carefully proved that Fe occupies Ti positions and not an other
vacant site. For that purpose, the X-ray powder diffraction pattern of
the iron-substituted compounds were compared to that from pure CoTiSb.
It was observed that no additional diffraction reflexes appear up to
10\% Fe-doping, thus confirming that the structure remains the same as
of pure CoTiSb, that is C1$_b$. The diffraction pattern of the pure and
a 10\% Fe doped sample are compared in Fig.\ref{fig-1}. The lattice
parameter $a = 0.5883(4)$nm for CoTiSb changes slightly (0.1\%) if 10\%
of Ti is replaced by Fe. The R values for the best fit in the Rietveld
refinement are R$_i$ = 6\% and R$_p$ = 11\% confirming the high degree
of site order of the sample.

\subsection{Magnetic properties}

The magnetic properties were investigated by means of magnetometry
(SQUID). To determine the saturation magnetisation hysteresis loops of
the CoTi$_{1-x}$Fe$_x$Sb samples for $x=0.01$ , 0.02, 0.05 and 0.1 were
measured. Already the sample with only 1\% Fe exhibit ferromagnetism at
room temperature. The total spin magnetic moment of the samples can be
predicted from the generalised Slater-Pauling rule \cite{Kue84} to be:
$m=N_V-18$, where $N_V$ is the number of valence electrons per unit
cell.

\begin{figure}[t]
\includegraphics[scale =1.0]{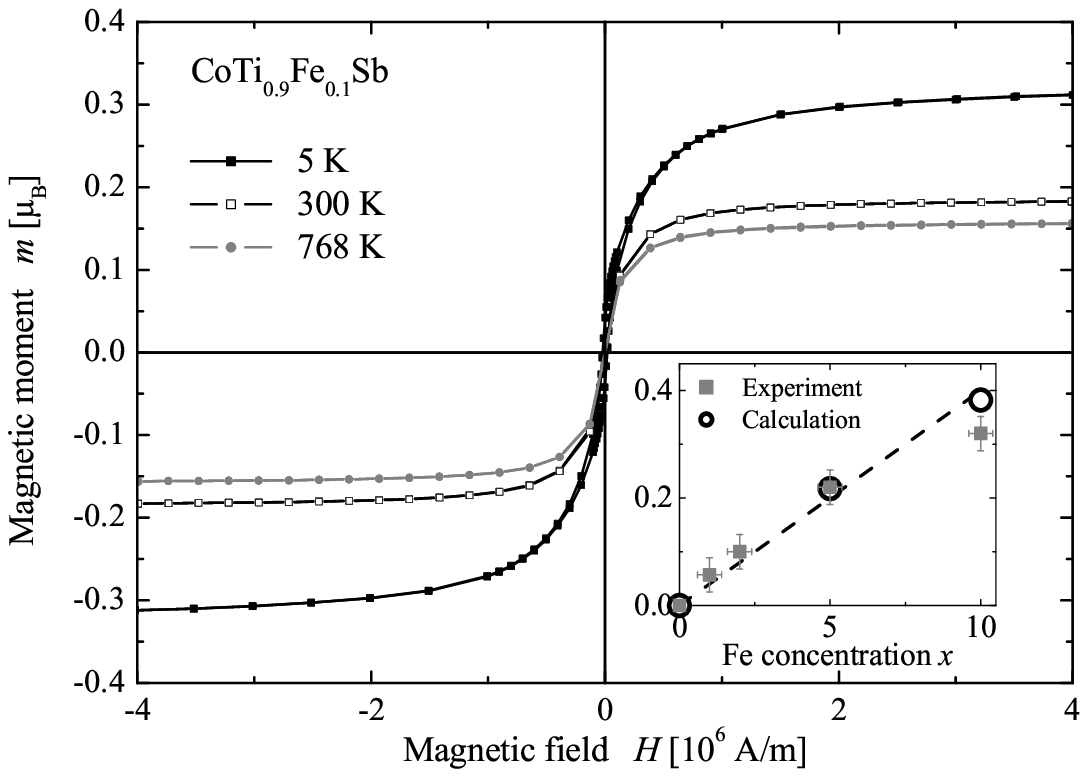}
\centering
\caption{Magnetic properties of CoTi$_{0.9}$Fe$_{0.1}$Sb. \newline
         The magnetisation as a function of the applied magnetic field
         at 5K, 300K and 768K is shown.
         Inset: The saturation magnetisation as a function von the Fe content is shown. 
         Compared are the experimental results with the calculated values.}
    \label{fig-2}
  \end{figure}

CoTiSb carries no magnetic moment according to the Slater-Pauling rule
because it has overall 18 valence electrons. Doping Fe on the Ti
positions should lead to a magnetic moment. In the case of
CoTi$_{0.9}$Fe$_{0.1}$Sb one can predict from the Slater-Pauling rule a
saturation magnetisation of 0.4~$\mu_{B}$ per formula unit due to the
18.4 valence electrons of the compound. In Fig.\ref{fig-2}. the
hysteresis loops of CoTi$_{0.9}$Fe$_{0.1}$Sb at 5K, 300K and 768K are
shown. Ferromagnetic behaviour is still observed at 768K which points
to a very high Curie temperature. The saturation magnetisation at 768K
has still half of the value at 5K. 
The calculated magnetic moments for the Fe-doped samples fit as well as 
the experimental values to the values one expected from 
the Slater-Pauling rule (see inset Fig.2).

For more details and other
properties of the series of CoTi$_{1-x}$Fe$_x$Sb with the Fe
concentration ranging from $x=0$ to 0.1 see Ref.\cite{KBF06}.

%%%%%%%%%%%%%%%%%%%% Figure 2 %%%%%%%%%%%%%%%%%%%%%%%%%%%%%%%%%%%%%%%%%%%%%%%%
%%%%%%%%%%%%%%%%%%%%%%%%%%%%%%%%%%%%%%%%%%%%%%%%%%%%%%%%%%%%%%%%%%%%%%%%%%%%%%

\section{Summary}

In summary, it was shown that Fe substituted \\
CoTi$_{1-x}$Fe$_x$Sb is
a new diluted semiconductor with a high Curie temperature of above
770K. Self consistent band structure calculations predict the alloy to
exhibit half-metallic ferromagnetism. The future idea is to design
devices which consist of alternating semi-conducting CoTiSb and
half-metallic ferromagnetic CoTi$_{1-x}$Fe$_x$Sb layers. In that case,
one may be able to design a device with the semi-conducting and
ferromagnetic parts being of almost the same material.

%in which & is replaced by \& including TEX command. So, when you see URL,
%please use the following address
%http://authors.elsevier.com/JournalDetail.html?PubID=505704&Precis=AIND

% The Appendices part is started with the command \appendix;
% appendix sections are then done as normal sections

\end{document}